\documentclass{article}

\usepackage{microtype}
\usepackage{graphicx}
\usepackage{subfigure}
\usepackage{booktabs} 
\usepackage{multirow}

\usepackage{hyperref}

\usepackage[accepted]{icml2019}

\icmltitlerunning{Representation Learning of Music Using Artist, Album, and Track Information}

\begin{document}

\twocolumn[
\icmltitle{Representation Learning of Music Using Artist, Album, and Track Information}

\begin{icmlauthorlist}
\icmlauthor{Jongpil Lee}{kaist}
\icmlauthor{Jiyoung Park}{naver}
\icmlauthor{Juhan Nam}{kaist}
\end{icmlauthorlist}

\icmlaffiliation{kaist}{Graduate School of Culture Technology, Korea Advanced Institute of Science and Technology (KAIST), Daejeon, South Korea}
\icmlaffiliation{naver}{NAVER Corp., South Korea}

\icmlcorrespondingauthor{Juhan Nam}{juhannam@kaist.ac.kr}

\icmlkeywords{Representation Learning, Artist label, Album label, Track label, Music Metadata}

\vskip 0.3in
]

\printAffiliationsAndNotice{}  

\begin{abstract}
Supervised music representation learning has been performed mainly using semantic labels such as music genres. However, annotating music with semantic labels requires time and cost. In this work, we investigate the use of factual metadata such as artist, album, and track information, which are naturally annotated to songs, for supervised music representation learning. The results show that each of the metadata has individual concept  characteristics, and using them jointly improves overall performance.
\end{abstract}

\section{Introduction}
\label{introduction}

Representation learning of music has been recently performed by supervised deep learning using semantic labels such as genres, moods and instruments \cite{choi2017transfer,lee2017multi}. However, annotating music with such semantic labels requires significant time and cost and the labels are often ambiguous, resulting in disagreement among annotators \cite{kim2017building}. Meanwhile, metadata such as artist labels require no cost and they are factual information with no ambiguity. We recently investigated the possibility of using artist information for representation learning of music and evaluated it in transfer learning settings \cite{park2017representation}. The results showed that the learned representation is comparable to those using the semantic labels. In this work, we extend the use of music metadata to album and track information, which are more specific levels than the artist information. We use a similarity-based learning model following the previous work and also report the effects of the number of negative samples and training samples.

\begin{figure}[t]
\vskip -0.1in
\begin{center}
\centerline{\includegraphics[width=0.85\columnwidth]{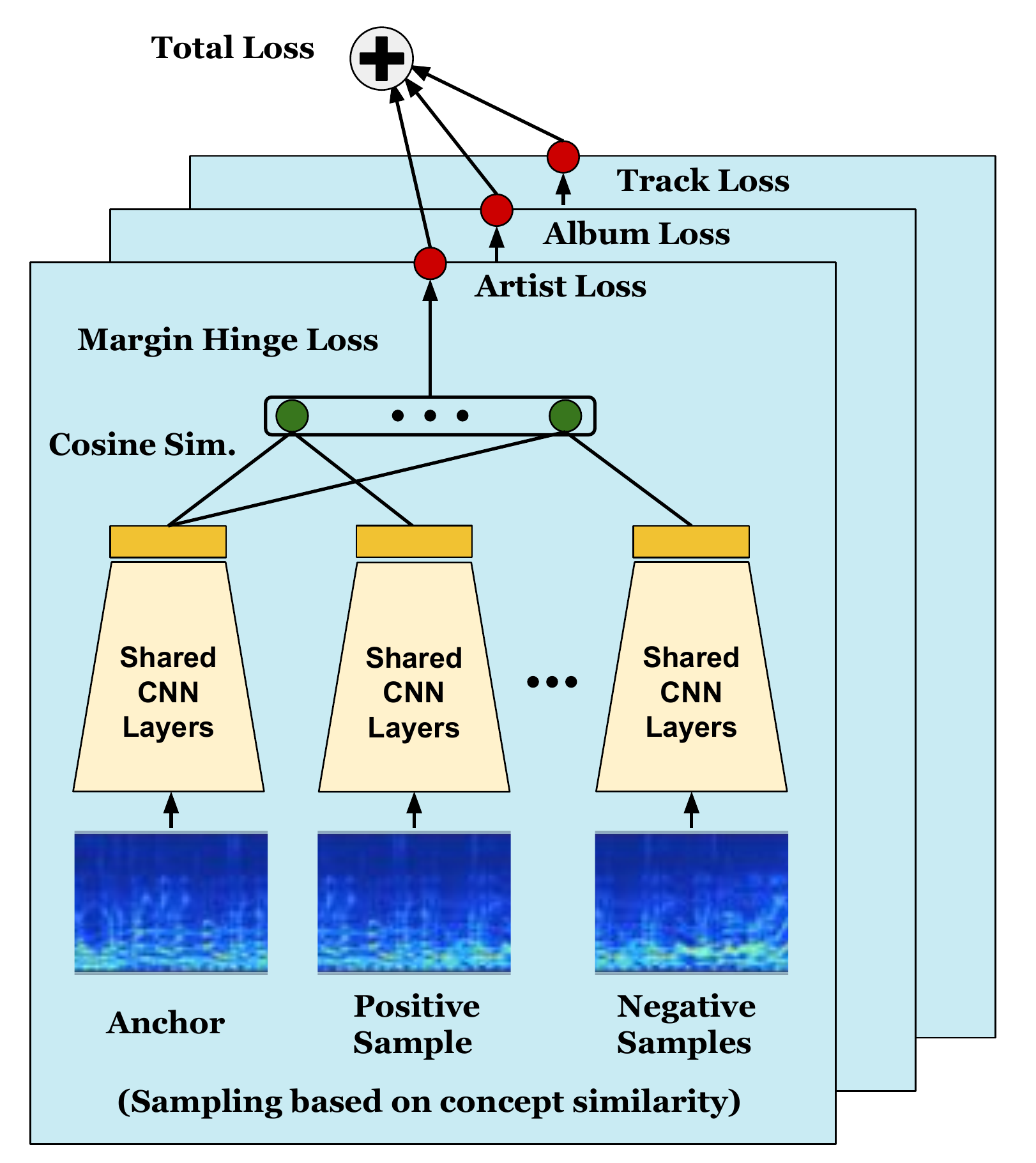}}
\caption{Joint learning model using artist, album, and track information.}
\label{mainfig}
\end{center}
\vskip -0.3in
\end{figure}

\section{Models}

Figure \ref{mainfig} illustrates the overview of representation learning model using artist, album, and track information. Following the previous work, we use a Siamese-style Convolutional Neural Network (CNN) with multiple negative samples\footnote{In this work, we used twice the number of filters for all layers.}. We build one large model that jointly learns artist, album, and track information and three single models that learns each of artist, album, and track information separately for comparison. The single model basically takes anchor sample, positive sample, and negative samples based on the similarity notion. For example, in the artist similarity concept, positive and negative samples are selected based on whether the sample is from the same artist as the anchor sample. We should note that the model takes a segment of audio (e.g. 3 second long), not the whole chunk of the song audio. Thus, in the track similarity concept, positive and negative samples are chosen based on whether the sample segment is from the same track as the anchor segment. Finally, we construct a joint learning model by simply adding three loss functions from the three similarity concepts, and share model parameters for all of them. 

 
\section{Experiments and Evaluations}

\begin{table}[t]
\caption{Hold-out positive and negative sample prediction.}
\label{holdout}
\vskip 0.10in
\begin{center}
\begin{small}
\begin{sc}

\resizebox{0.70\columnwidth}{!}{\begin{tabular}{@{}ccccc@{}}
\toprule
\begin{tabular}[c]{@{}c@{}} Learned\\ concept \end{tabular} & Artist & Album & Track & \begin{tabular}[c]{@{}c@{}}Artist\\ +Album\\ +Track\end{tabular} \\ \midrule
Artist & 0.680 & 0.634 & 0.539 & 0.686 \\
Album & 0.732 & 0.822 & 0.653 & 0.763 \\
Track & 0.922 & 0.958 & 0.971 & 0.945 \\ \bottomrule
\end{tabular}}

\end{sc}
\end{small}
\end{center}
\vskip -0.1in
\end{table}

\begin{table}[t]
\caption{Transfer learning experiment. Baseline results are generated by performing genre classification directly without transfer learning.
}
\label{transfer}
\begin{center}
\begin{small}
\begin{sc}

\resizebox{\columnwidth}{!}{\begin{tabular}{@{}cccccc@{}}
\toprule
\begin{tabular}[c]{@{}c@{}} Learned\\ concept \end{tabular} & \begin{tabular}[c]{@{}c@{}} genres \\ (baseline) \end{tabular} & Artist & Album & Track & \begin{tabular}[c]{@{}c@{}}Artist\\ +Album\\ +Track\end{tabular} \\ \midrule
GTZAN & 0.547 & 0.724 & 0.652 & 0.564 & 0.745 \\
FMA small & 0.533 & 0.598 & 0.560 & 0.463 & 0.593 \\
NAVER Korean & 0.720 & 0.662 & 0.641 & 0.549 & 0.663 \\ 
\bottomrule
\end{tabular}}

\end{sc}
\end{small}
\end{center}
\vskip -0.1in
\end{table}

\begin{table}[t]
\caption{The effect of the number of negative samples. The model is trained with 1000 artists, 2000 albums, and its related track concepts.}
\label{negsamples}
\begin{center}
\begin{small}
\begin{sc}

\resizebox{0.9\columnwidth}{!}{\begin{tabular}{@{}cccccc@{}}
\toprule
\begin{tabular}[c]{@{}c@{}}Number of \\ Negative Samples \end{tabular}  & 1 & 2 & 4 & 8 & 16 \\ \midrule
GTZAN & 0.665 & 0.663 & 0.681 & 0.702 & 0.711 \\
FMA small & 0.544 & 0.535 & 0.568 & 0.573 & 0.578 \\
NAVER Korean & 0.643 & 0.634 & 0.658 & 0.676 & 0.673 \\ \bottomrule
\end{tabular}}

\end{sc}
\end{small}
\end{center}
\vskip -0.1in
\end{table}

\begin{table}[t]
\caption{The effect of the number of training samples. The model is trained with 4 negative samples with artist, album, and track concepts. The number of albums used is twice the number of artists.}
\label{numbersongs}
\begin{center}
\begin{small}
\begin{sc}

\resizebox{0.9\columnwidth}{!}{\begin{tabular}{@{}cccccc@{}}
\toprule
\begin{tabular}[c]{@{}c@{}}Number of \\ Training Artists \end{tabular}  & 500 & 1000 & 2000 & 5000 & 10000 \\ \midrule
GTZAN & 0.638 & 0.681 & 0.706 & 0.745 & 0.755 \\
FMA small & 0.517 & 0.568 & 0.588 & 0.593 & 0.603 \\
NAVER Korean & 0.636 & 0.658 & 0.668 & 0.663 & 0.686 \\ \bottomrule
\end{tabular}}

\end{sc}
\end{small}
\end{center}
\vskip -0.1in
\end{table}

The four models are trained with Million Song Dataset (MSD) and its artist and album metadata \cite{bertin2011million}. We first build two splits based on each artist and album information. The artist split is the same as the previous work, which has 20 songs for each artist. For the album split, we selected 10 songs for each album and used twice as many albums to match the number of training samples of artist. Then, 10 songs of one album are divided into 8 songs, 1 song, and 1 song for training, validation and testing. The artist split is twice these numbers. For the validation sampling of artist or album concept, the positive sample is selected from the training set and the negative samples are chosen from the validation set based on the validation anchor's concept. For the track concept, it basically follows the artist split, and the positive sample for the validation sampling is chosen from the other part of the anchor song. 

The evaluation is conducted in two ways: 1) hold-out positive and negative sample prediction and 2) transfer learning experiment. The hold-out positive and negative sample prediction was designed to see how well the models distinguish each concept. The evaluation is conducted on the test set of the above splits. For the artist and album concept, the positive sample is selected from the validation set and the negative samples are from test set based on the anchor's concept. In this evaluation, the random guess is 20\% when the model uses 4 negative samples. The transfer learning experiment is performed on three external genre classification datasets including GTZAN (a fault-filtered version) \cite{tzanetakis2002musical,kereliuk2015deep}, FMA small \cite{fma_dataset}, and NAVER Korean \cite{park2017representation}. In this experiment, the learned representation is extracted and injected into a linear softmax classifier. This experiment was designed to see the generalization ability of the learned representations. For both evaluations, we used a model trained with 5000 artists (or/and 10000 albums) with 4 negative samples. After grid search, the margin values of loss function were set to 0.4, 0.25, and 0.1 for artist, album, and track concepts, respectively.

\section{Results}
The result of hold-out positive and negative sample prediction is shown in Table \ref{holdout}. We can see that each of the models performs best when the concept matches between the training set and test set. Also, the jointly learned model achieves good performance for all concepts.

The transfer learning experiment result is shown in Table \ref{transfer}. The artist model shows the best performance among the three single concept models, followed by the album model. This is probably because the genre classification task is more similar to the artist concept discrimination than album or track. The jointly learned model slightly outperforms the artist model. Finally, we included the baseline results obtained by performing genre classification directly without transfer learning. The results show that transfer learning using large music corpora with the factual metadata is highly effective in the GTZAN and FMA datasets, but not in NAVER dataset. This was due to the cross-cultural differences between the source and target datasets when looking closely at class-wise performances. 

The effects of the number of negative samples and the number of training samples are shown in Table \ref{negsamples} and Table \ref{numbersongs}, respectively. We can see that increasing the number of negative samples and the number of training songs improves the model performance as expected.


\bibliography{example_paper}

\begin{thebibliography}{8}
\providecommand{\natexlab}[1]{#1}
\providecommand{\url}[1]{\texttt{#1}}
\expandafter\ifx\csname urlstyle\endcsname\relax
  \providecommand{\doi}[1]{doi: #1}\else
  \providecommand{\doi}{doi: \begingroup \urlstyle{rm}\Url}\fi

\bibitem[Bertin-Mahieux et~al.(2011)Bertin-Mahieux, Ellis, Whitman, and
  Lamere]{bertin2011million}
Bertin-Mahieux, T., Ellis, D.~P., Whitman, B., and Lamere, P.
\newblock The million song dataset.
\newblock In \emph{Proc. of the International Society for Music Information
  Retrieval Conference (ISMIR)}, volume~2, pp.\  591--596, 2011.

\bibitem[Choi et~al.(2017)Choi, Fazekas, Sandler, and Cho]{choi2017transfer}
Choi, K., Fazekas, G., Sandler, M., and Cho, K.
\newblock Transfer learning for music classification and regression tasks.
\newblock In \emph{Proc. International Society for Music Information Retrieval
  Conf.}, pp.\  141--149, 2017.

\bibitem[Defferrard et~al.(2017)Defferrard, Benzi, Vandergheynst, and
  Bresson]{fma_dataset}
Defferrard, M., Benzi, K., Vandergheynst, P., and Bresson, X.
\newblock Fma: A dataset for music analysis.
\newblock In \emph{Proc. of the International Society for Music Information
  Retrieval Conference (ISMIR)}, pp.\  316--323, 2017.

\bibitem[Kereliuk et~al.(2015)Kereliuk, Sturm, and Larsen]{kereliuk2015deep}
Kereliuk, C., Sturm, B.~L., and Larsen, J.
\newblock Deep learning and music adversaries.
\newblock \emph{IEEE Transactions on Multimedia}, 17\penalty0 (11):\penalty0
  2059--2071, 2015.

\bibitem[Kim et~al.(2017)Kim, Kum, Park, Lee, Park, and Nam]{kim2017building}
Kim, K.~L., Kum, S., Park, C.~L., Lee, J., Park, J., and Nam, J.
\newblock Building k-pop singing voice tag dataset: A progress report.
\newblock In \emph{Late Breaking Demo in the International Society for Music
  Information Retrieval Conf.}, 2017.

\bibitem[Lee \& Nam(2017)Lee and Nam]{lee2017multi}
Lee, J. and Nam, J.
\newblock Multi-level and multi-scale feature aggregation using pretrained
  convolutional neural networks for music auto-tagging.
\newblock \emph{IEEE signal processing letters}, 24\penalty0 (8):\penalty0
  1208--1212, 2017.

\bibitem[Park et~al.(2018)Park, Lee, Park, Ha, and Nam]{park2017representation}
Park, J., Lee, J., Park, J., Ha, J., and Nam, J.
\newblock Representation learning of music using artist labels.
\newblock In \emph{Proc. of International Society for Music Information
  Retrieval Conference (ISMIR)}, pp.\  717--724, 2018.

\bibitem[Tzanetakis \& Cook(2002)Tzanetakis and Cook]{tzanetakis2002musical}
Tzanetakis, G. and Cook, P.
\newblock Musical genre classification of audio signals.
\newblock \emph{IEEE Transactions on speech and audio processing}, 10\penalty0
  (5):\penalty0 293--302, 2002.

\end{thebibliography}
\bibliographystyle{icml2019}

\end{document}